\documentclass[aip,amsmath,amssymb,reprint]{revtex4-1}

\usepackage{graphicx}
\usepackage{dcolumn}
\usepackage{bm}
\usepackage{comment}
\usepackage[utf8]{inputenc}
\usepackage[T1]{fontenc}
\usepackage{mathptmx}
\usepackage{etoolbox}

\makeatletter
\def\@email#1#2{%
 \endgroup
 \patchcmd{\titleblock@produce}
  {\frontmatter@RRAPformat}
  {\frontmatter@RRAPformat{\produce@RRAP{*#1\href{mailto:#2}{#2}}}\frontmatter@RRAPformat}
  {}{}
}%
\makeatother
\begin{document}

\preprint{AIP/123-QED}

\title{Spatial Dependence of Microscopic Percolation Conduction}
\author{Matthew D. Golden}
 \altaffiliation[Now at ]{Department of Physics and Astronomy, Purdue University, West Lafayette, IN 47907.}
\author{Joseph P. Straley}
 \email{jstraley@uky.edu}
\affiliation{ 
Department of Physics \& Astronomy \\University of Kentucky \\Lexington, KY 40506
}

\date{\today}

\begin{abstract}
In two dimensions, the average electrical conductance from a point in a percolating network to the network boundary should be related by a conformal transformation to the conductance from one point to another in an unbounded network.  We verify that this works at the percolation threshold for the square.
\end{abstract}

\maketitle

\section{Microscopic percolation}

The percolation model studies the properties of networks derived from a lattice by random deletions of bonds.   It comprises several different questions.  The earliest studies established that as the fraction $p$ of bonds that are present is increased, there is a kind of phase transition between networks containing only finite connected clusters and networks that contain an "infinite" connected cluster (or one that spans a large finite network from side to side)\cite{BH}.  In two dimensions this "connectedness" model can be understood as the limiting case Q $\rightarrow$ 1 of the Q-component Potts models; the exponents characterizing the correlation length and probability of a lattice site belonging to the infinite cluster are given by the den Nijs\cite{dennijs, nienhuis, pearson} relationships that describes all exponents of Potts models. 
Relevant to this study are the correlation length $\xi$
\begin{equation}
\xi \approx |p - p_c|^{- \nu }   ,
\end{equation}
and the probability $P$ that an arbitrary site belongs to the infinite cluster    
\begin{equation} P \approx   |p-p_c|^\beta
\end{equation}
where $\nu = 4/3$  and $\beta = 5/36$. 
These basic properties of a percolating system can be readily studied in computer simulations.

Connectedness isn't a very useful property of a real-world experimental system.   This has led to the study of the properties of a percolating system that can be measured by established experimental methods.   A paradigm model is to assign unit conductance to the bonds and measure the "macroscopic" electrical conductivity: the average current density produced by a uniform electric field\cite{stral1}.   In two dimensions the macroscopic electrical conductance is well described for $p > p_c$  by
\begin{equation}
G =G_0  (p - p_c)^t
\end{equation}
and at $p = p_c$ by
\begin{equation}
G = G_0  L^{-t/\nu}
\end{equation}
Simulations\cite{lobb}  indicate  $t/\nu = 0.973$, independent of details of the model (e.g. periodic boundary conditions vs. assigned potentials at the boundaries, as well as the details how the sites are connected).  The relationship between this exponent and those for the connectedness problem is unclear; unusual for a two dimensional model, the exponent does not appear to be a rational number.  

 Another model problem, the subject of the present paper, is the "microscopic" conductance: the conductance from a small source to a small sink at distance D, or the current from a small source to a grounded boundary at distance D/2.  This seems to be a simpler and more prospective model to study than the macroscopic conductance, because for a chosen boundary shape it depends on just the distance $D$.  

In studying the microscopic conductance we found it useful to ignore configurations for which the chosen site does not have a connected path to the boundary, and only average the conductance over the remainder, because the variance of the distribution of conductances is smaller than the variance of connectedness itself  (the distribution of connectedness is binary), and the connectedness problem is well studied.   Thus for conductance from one point to another at distance $D$ in an infinite system we define
\begin{equation}
g_{connected} =g_0 D^{-u}
\end{equation}
The conductance defined in terms of all configurations would include the statistics of connectedness, so that
\begin{equation}
g_{all} = D^{-\beta/\nu}   g_{connected}  .
\end{equation}
Conductance from one lattice site to a boundary is a well-defined concept.   However, in going to a continuum version of the problem we find that the current field is singular at the insertion point, so that the conductance seems to be zero.   This difficulty in the continuum limit can be avoided by making the insertion a small disk.  The disk size sets the scale factor for distance and slightly complicates the meaning of "conductance from a point."
\section{Conformal invariance}

Theoretical understanding of the percolation conduction models has been hindered by a lack of understanding how to incorporate these effects into the theoretical framework, since what is measured is not the derivative of a free energy with respect to an applied field, nor the correlation function of known operators.   As a step towards gaining a better understanding of these problems we studied how the microscopic conductance from an interior point to the boundary depends on the position of the interior point.

At the critical point, two-dimensional critical models often have conformal invariance\cite{cardy}, which is a symmetry that contains translational, rotational, and scale invariance.  Representing the (x,y) coordinates of a point by the complex number z = x + i y,   all analytic functions give conformal transformations.    This implies a relationship between the critical behavior of correlation functions for systems of different shape.  Relevant to our study, the square $[-L/2 < x < L/2, 0 < y < L]$ is mapped into the upper half plane by the Jacobian elliptic functions.   We used them in the form
\begin{equation}
w =  sn(2Kz/L,m) 
\end{equation}
where $z = x + iy$ and $w$ is the corresponding coordinate in the upper half plane, $m$  is the modulus of the elliptic function (it takes the value $0.0294376$ for the proposed mapping) and $K$ is the quarter period of the elliptic function ($K = 1.58255$ and $K' = 2 K$, for the mapping we will use).  The boundary of the square $-L/2 < x < L/2$, $0 < y < L$ is mapped onto the real axis.   The conductance from a point $w$ to the real axis can only depend on $Im \  w$ as a power law (at the percolation threshold).  Treating this like a correlation function for a spinless field, conformal field theory implies the conductance to the boundary of a rectangle has the form $g(z) = (D/L)^{-u}$, where the "distance function" D has the form\cite{res}
\begin{equation}
D/L = \frac{ Im \  sn(2Kz/L,m)}{Re \ |cn(2Kz/L,m) dn(2Kz/L,m)|} 
\label{Dfun}
\end{equation}
 
Despite the asymmetrical form, this distance function has the symmetry of the square; it smoothly interpolates between being the distance to the nearest part of the boundary, for points near the boundary.

We calculated the microscopic conductance for a source point in the square in a computer simulation.  For a chosen realization and chosen interior point, we first verified that the interior point belongs to a cluster that touches the boundary.   Kirchhoff's equations give relationships between the voltages of neighboring points; these were simplified by removing links that could not carry a current, and the result was solved by a pivoted Gaussian elimination. For further details concerning the simulation, see the Appendix. 

When the current is injected at the center of a square of edge $L$, the probability that the site is connected to the boundary is expended to vary as $L^{-\tilde \beta}$ where $\tilde \beta = \beta/\nu = 5/48 = 0.1048$, and the average conductance of the connected configurations should vary as $L^{-\tilde t}$, with $\tilde t = t/\nu = 0.972$. We made 30,000 random realizations of the system for $L = 30,40, 60, 80, 120, 160,$ and  $180 $ and averaged the resulting conductance from the center to the edge, finding $\tilde \beta  = 0.106973$ and $\tilde t = 1.00276$; these are not too different from the expected values.

For points not at the center, we found that the rule $G = G_0 D^{-u}$ (using the distance function $D$ given in Eq.\ref{Dfun}) works reasonably well throughout the square, as shown by Figures (1) and (2), which compare the measured (black) and calculated (red)conductance for $L = 180$ and  $L = 100$ along lines of constant $Y$.   The best fit for $L = 180$ is obtained for $u = 1.08$, and for $L = 100$, $u = 1.12$.  We believe the disagreements between the various estimates are due to the finite lattice spacing, since the discrepancy decreases with lattice size, and increases towards the edges of the sample.

\begin{figure}
\includegraphics[scale=.5]
{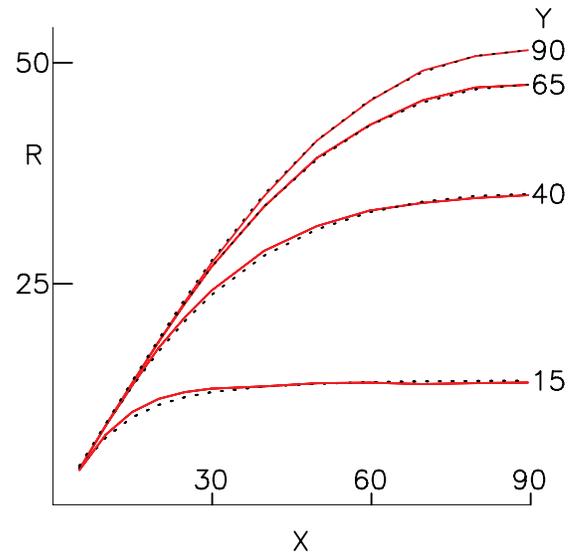}
\caption{\label{fig:L=100a} Comparison numerical experiment  (red) and theory (black dots) for L=100.  The curves represent the reciprocal of the average conductance of the connected configurations for various $0 < X < 50$ at fixed Y, for Y = 5, 10, 15, 20, 25, 30, 40, and 50.  The scaling exponent is 1.12}
\end{figure}

\begin{figure}
\includegraphics[scale=.5]{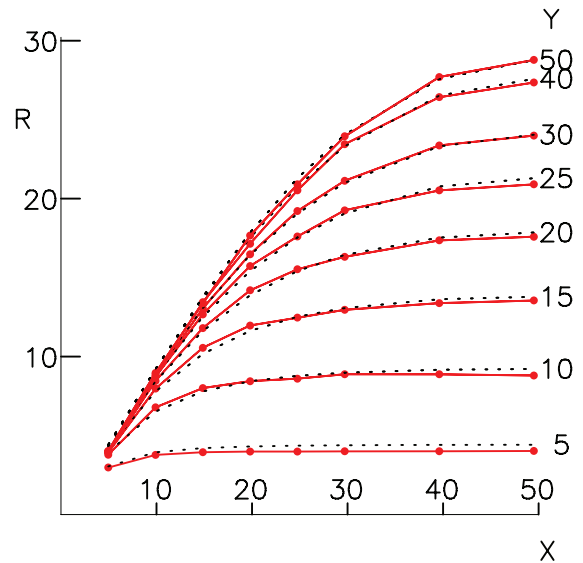}
\caption{\label{fig:L=180a}Comparison of numerical experiment (red) and theory (black dots) for L=180. The curves represent the reciprocal of the average conductance of the connected configurations for various $0 < X < 90$ at fixed Y, for = 15, 40, 65, and 90. The scaling exponent is 1.08 }
\end{figure}

\section{conclusion}
The agreement between the representation of the dependence of the conductance on the distance D (\ref{Dfun}) and the results of simulations shown in the Figures  supports the idea that the microscopic conductance can be regarded to be a kind of correlation function.

\begin{acknowledgments}
 M.D.G. acknowledges support by the University of Kentucky Research
Experiences for Undergraduates program, funded by the NSF award PHY-1950795.

We would thank the University of Kentucky Center for Computational Sciences and Information Technology Services Research Computing for their support and use of the Lipscomb Compute Cluster and associated research computing resources.

\end{acknowledgments}

\section*{Data Availability Statement}

 See the Supplemental Material at [URL will be inserted by publisher] for the numerical values used in constructing the figures in this paper.

\appendix

\section{Supplemental materials} 
The simulations were made with square samples of bond percolating networks at $p_c = 1/2$.  The samples contained LxL interior sites.  For general $X$ and $Y$ the edges (X,0),(0,Y), (X,L+1), (L+1,Y) were held a zero potential (so that the center site is (L/2,L/2); unit potential was applied at a chosen site $(X,Y)$ and the current to the boundary was calculated, many times.  The resulting conductance was averaged over the subset of the realizations that had nonzero conductance.  The fraction of conducting realizations is also noted.
\begin{table}
\caption{\label{tab:table1} The case $L = 180$. More than 20,000 realizations were made for each position.   For all cases the standard error of the mean was less than 0.5\% of the average. Column 1: Position X; Column 2: Position Y; Column 3: Fraction of samples that were conducting; Column 4: Conductance averaged over all samples; Column 5: Conductance averaged over samples with nonzero conductance }
\begin{ruledtabular}
\begin{tabular}{lllll}
X&Y& Fraction & Conductance & Conductance \\
\hline
 5 & 90 & 0.769 & 0.1916 & 0.2491 \\
10 & 90 & 0.708 & 0.0786 & 0.1111 \\
15 & 90 & 0.679 & 0.0491 & 0.0724 \\
20 & 90 & 0.654 & 0.0354 & 0.0541 \\
25 & 90 & 0.644 & 0.0280 & 0.0435 \\
30 & 90 & 0.632 & 0.0231 & 0.0366 \\
40 & 90 & 0.617 & 0.0177 & 0.0286 \\
50 & 90 & 0.606 & 0.0147 & 0.0243 \\
60 & 90 & 0.602 & 0.0132 & 0.0219 \\
70 & 90 & 0.599 & 0.0122 & 0.0204 \\
80 & 90 & 0.594 & 0.0117 & 0.0197 \\
90 & 90 & 0.593 & 0.0115 & 0.0194 \\
 5 & 40 & 0.768 & 0.1921 & 0.2501 \\
10 & 40 & 0.711 & 0.0796 & 0.1120 \\
15 & 40 & 0.678 & 0.0503 & 0.0741 \\
20 & 40 & 0.663 & 0.0374 & 0.0564 \\
25 & 40 & 0.653 & 0.0307 & 0.0471 \\
30 & 40 & 0.641 & 0.0265 & 0.0413 \\
40 & 40 & 0.632 & 0.0220 & 0.0348 \\
50 & 40 & 0.623 & 0.0198 & 0.0317 \\
60 & 40 & 0.618 & 0.0186 & 0.0300 \\
70 & 40 & 0.616 & 0.0180 & 0.0293 \\
80 & 40 & 0.617 & 0.0178 & 0.0288 \\
90 & 40 & 0.620 & 0.0177 & 0.0285 \\
 5 & 65 & 0.769 & 0.1920 & 0.2497 \\
10 & 65 & 0.708 & 0.0789 & 0.1114 \\
15 & 65 & 0.680 & 0.0495 & 0.0728 \\
20 & 65 & 0.660 & 0.0358 & 0.0543 \\
25 & 65 & 0.646 & 0.0284 & 0.0439 \\
30 & 65 & 0.633 & 0.0235 & 0.0372 \\
40 & 65 & 0.620 & 0.0184 & 0.0296 \\
50 & 65 & 0.611 & 0.0156 & 0.0255 \\
60 & 65 & 0.601 & 0.0140 & 0.0233 \\
70 & 65 & 0.607 & 0.0133 & 0.0219 \\
80 & 65 & 0.601 & 0.0127 & 0.0212 \\
90 & 65 & 0.597 & 0.0126 & 0.0210 \\
 5 & 15 & 0.773 & 0.1969 & 0.2547 \\
10 & 15 & 0.724 & 0.0912 & 0.1261 \\
15 & 15 & 0.701 & 0.0668 & 0.0953 \\
20 & 15 & 0.698 & 0.0583 & 0.0835 \\
25 & 15 & 0.678 & 0.0532 & 0.0785 \\
30 & 15 & 0.683 & 0.0521 & 0.0762 \\
40 & 15 & 0.683 & 0.0510 & 0.0747 \\
50 & 15 & 0.675 & 0.0491 & 0.0727 \\
60 & 15 & 0.676 & 0.0489 & 0.0724 \\
70 & 15 & 0.673 & 0.0493 & 0.0733 \\
80 & 15 & 0.679 & 0.0493 & 0.0727 \\
90 & 15 & 0.674 & 0.0488 & 0.0724 \\
\end{tabular}
\end{ruledtabular}
\end{table}

\begin{table}
\caption{\label{tab:table2} The case $L = 100$.  50,000 realizations were made for each position. For all cases the standard error of the mean was less than 0.6\% of the average. Column 1: Position X; Column 2: Position Y; Column 3: Fraction of samples that were conducting; Column 4: Conductance averaged over all samples; Column 5: Conductance averaged over samples with nonzero conductance }
\begin{ruledtabular}
\begin{tabular}{lllll}
X&Y& Fraction & Conductance & Conductance \\
\hline
 5 & 10 & 0.770 & 0.2029 & 0.2635 \\ 
 5 & 15 & 0.767 & 0.1950 & 0.2540 \\ 
10 & 15 & 0.717 & 0.0900 & 0.1254 \\ 
 5 & 20 & 0.777 & 0.1969 & 0.2534 \\ 
10 & 20 & 0.711 & 0.0840 & 0.1181 \\ 
15 & 20 & 0.691 & 0.0585 & 0.0847 \\ 
 5 & 25 & 0.778 & 0.1950 & 0.2508 \\ 
10 & 25 & 0.711 & 0.0816 & 0.1146 \\ 
15 & 25 & 0.687 & 0.0542 & 0.0789 \\ 
20 & 25 & 0.675 & 0.0429 & 0.0635 \\ 
 5 & 30 & 0.774 & 0.1959 & 0.2531 \\ 
10 & 30 & 0.709 & 0.0811 & 0.1143 \\ 
15 & 30 & 0.685 & 0.0524 & 0.0766 \\ 
20 & 30 & 0.665 & 0.0403 & 0.0607 \\ 
25 & 30 & 0.653 & 0.0340 & 0.0520 \\ 
 5 & 40 & 0.764 & 0.1916 & 0.2507 \\ 
10 & 40 & 0.707 & 0.0798 & 0.1127 \\ 
15 & 40 & 0.684 & 0.0509 & 0.0744 \\ 
20 & 40 & 0.659 & 0.0384 & 0.0583 \\ 
25 & 40 & 0.653 & 0.0318 & 0.0487 \\ 
30 & 40 & 0.654 & 0.0279 & 0.0426 \\ 
 5 & 50 & 0.774 & 0.1918 & 0.2478 \\ 
10 & 50 & 0.712 & 0.0795 & 0.1115 \\ 
15 & 50 & 0.677 & 0.0506 & 0.0747 \\ 
20 & 50 & 0.657 & 0.0372 & 0.0566 \\ 
25 & 50 & 0.642 & 0.0307 & 0.0478 \\ 
30 & 50 & 0.642 & 0.0268 & 0.0418 \\ 
40 & 50 & 0.642 & 0.0232 & 0.0361 \\

\end{tabular}
\end{ruledtabular}

\end{table}

\end{document}